\documentclass[prl,aps,groupedaddress,twocolumn]{revtex4}

\usepackage{xspace,amsmath,amsfonts,amsthm,amssymb,amsbsy}
\usepackage{graphicx}
\usepackage{amssymb}
\usepackage{color}
\usepackage{psfrag}
\usepackage{ifsym}
\usepackage{epstopdf}
\usepackage{units}

\usepackage{epigraph}

\usepackage[breaklinks=true]{hyperref}
\hypersetup{
  colorlinks   = true, 
  urlcolor     = blue, 
  linkcolor    = cyan, 
  citecolor   =  red 
}

\newcommand{\Int}{\int\limits}

\definecolor{pink}{rgb}{1,0.078,0.57}

\definecolor{green}{rgb}{0,0.7,0.9}

\newcommand{\ket}[2] {| #1 \rangle_{#2}}
\newcommand{\bra}[2] {\langle #1 |_{#2}}
\newcommand{\braket}[2] {\langle #1 | #2 \rangle}

\newcommand{\bE}{\mathbf{E}}

\newcommand{\hc}{\mathrm{h.c.}}

\newcommand{\uv}[1]{\hat{\mathbf{#1}}}

\newcommand{\n}{^{(\gamma)}}
\newcommand{\lavg}{\left \langle}
\newcommand{\ravg}{\right \rangle_{\Gamma}}

\begin{document}

\title{Transform-Limited-Pulse Representation of Excitation \\ with Natural Incoherent Light}

\author{Aur\'elia Chenu}
\email[ Corresponding author: ]{aurelia.chenu@utoronto.ca}
\author{Paul Brumer}

\affiliation{Chemical Physics Theory Group and Centre for Quantum Information and Quantum Control, University of Toronto, 80 Saint George Street, Toronto, Ontario, M5S 3H6 Canada}


\begin{abstract}
 We study the natural excitation of molecular systems, applicable to, for example, photosynthetic light-harvesting complexes, by natural incoherent light.  In contrast with the conventional classical models, we show that the light need not have random character to properly represent the resultant linear excitation. Rather, thermal excitation can be interpreted as a collection of individual events resulting from the system's interaction with individual, deterministic pulsed realizations that  constitute the field. The derived expressions for the  individual field realizations and excitation events allow for a wave function formalism, and therefore constitute a useful calculational tool to study dynamics following thermal-light excitation. Further, they provide a route to the experimental determination of natural incoherent excitation using pulsed laser techniques.
\end{abstract}

\maketitle

\epigraph{``\textit{One of the most important attributes of a stationary random process is its spectrum.}''}{L. Mandel \& E. Wolf \cite{MandelWolfBook2}}

The process of light-matter interaction can initiate fundamental dynamics in systems as diverse as semiconductors, quantum dots, and living organisms. Of particular interest is the linear interaction of sunlight with molecules such as photosynthetic light-harvesting complexes \cite{BlankenshipBook}, which is the foremost step responsible for primary energy production on earth.

While the development of advanced experiments such as non-linear spectroscopy \cite{Grondelle2006a} allows characterization of the system's dynamical properties upon photo-excitation, the measured molecular dynamics can differ from that resulting from natural excitation conditions \cite{Jiang1991a, Mancal2010a, Hoki2011a, Brumer2012a, Kassal2013a, Han2013a}, mainly because of significant differences between the exciting lights \cite{Chenu2014b}. Although in theory, it is  possible to measure molecular dynamics directly from incoherent excitation \cite{Turner2013a}, technical difficulties are such that, in practice, no experimental data are yet available. 
At present, the dynamics of a system excited by thermal light can only be inferred from appropriate theoretical models, which require a proper description of thermal excitation. Two different approaches are available: (i) a full quantum-mechanical treatment, using a quantum description of the light (well-known for thermal light, e.g.  the $P$-representation, representing the light state by direct products of incoherent mixtures of coherent states associated with a single frequency---typically plane waves \cite{Loudon2000}), 
or (ii) a semi-classical approach---based on a classical description of the radiation field.

Thermal light can only be represented statistically by a probability distribution function or, alternatively, as a set of all possible realizations. 
The most conventional classical representation -- collision-broadening model \cite{Loudon2000} --  relies on realizations with stochastic phase jumps to reproduce the random character of thermal light, with the average time interval between jumps reproducing the (short) coherence time. 
Here we show that no random parameter is needed for the proper representation of \emph{linear} excitation, i.e. weak excitation, and that, in first-order perturbation theory, an ensemble of \emph{deterministic} realizations can reproduce the result of thermal-light excitation. 
Indeed, even pulses are possible. In studying the interaction of a molecule with an ensemble of fields (representing the individual realizations), 
we provide answers to the following question: 
What field characteristics provide a proper description such that the ensemble of individual realizations yield the correct molecular density matrix? 

We address this question within the framework of a molecule, applicable to, e.g., photosynthetic pigment-protein complexes such as the Fenna-Matthew-Olson (FMO) complex \cite{Adolphs2006a}, or the LH2 light-harvesting complex \cite{Cogdell2006a}. 
Because we focus on the molecular response to the light,  
the relevance and interpretation of our results are independent of the effects of the protein environment, and we assume a closed molecular system. 

Consider the Hamiltonian
\begin{equation}\label{eq:H}
H(t) = \sum_{\alpha} \varepsilon_\alpha \ket{\alpha}{} \bra{\alpha}{}   -\bold{d} \cdot \left(\bE\n(t)+\bE^{(\gamma)}(t)^*\right),
\end{equation}
where the system Hamiltonian  (first term on the r.h.s.) is written in the basis
of eigenstates $\ket{\{\alpha\}}{}$ with eigenenergies $\varepsilon_\alpha$, and
the second term describes the interaction with the field in the semi-classical approach. Here
$\bold{d} = \sum_\alpha \mu_\alpha  \ket{\alpha}{}\bra{g}{} \uv{n}_\alpha+ \hc$, with $\ket{g}{}$ the molecular ground state, 
is the transition dipole moment operator,  oriented along the unit vector $\uv{n}_\alpha$;
$\bE\n(t)$ denotes the positive frequency part of one individual field realization.
The superscript $\gamma$ is used throughout the paper to label a particular realization,
either of the field, or of the corresponding molecular excited state. 
This Hamiltonian characterizes the molecular dynamics  upon interaction with one of the field realizations.

In the rotating wave approximation and in the limit of weak-field interaction,
first-order perturbation theory provides a solution for 
 the wave function  resulting from the $\ket{g}{}$ state excitation as a superposition of eigenstates: 
\begin{equation} \label{eq:psi_alpha_def}
\ket{\psi\n(t)}{}  = - \frac{i}{\hbar} \sum_{\alpha=1}^{N_\alpha} \mu_\alpha e^{-i \omega_\alpha t}  \int_{0}^t  d\tau \, e^{i \omega_\alpha \tau} E\n(\tau) \ket{\alpha}{},
\end{equation}
where $N_\alpha$ is the number of molecular eigenstates and $\omega_\alpha$ denotes the eigenfrequencies. 
 Here, we assume the excited-state level splitting to be larger than their spontaneous decay widths
and neglect effects from radiative decay  \cite{Tscherbul2014b}. 
For the sake of clarity, we  consider a one-dimensional field, and assume it lies along the orientation of the transition dipole moment $\uv{n}_\alpha$, taken identical for all eigenstates.

Considering that the molecule's response to  a stochastic field is composed by an ensemble of  individual excitation events, we define the density matrix: 
\begin{equation}\label{eq:rho_tot_def}
\rho^{\rm tot}(t) \equiv \lavg \ket{\psi\n(t)}{} \bra{\psi\n(t)}{} \ravg , 
\end{equation}
where $\lavg \dots \ravg$ denotes   averaging over the ensemble of realizations. 
From (\ref{eq:psi_alpha_def}), the elements of this density matrix are:
\begin{equation} \label{eq:rho_tot}
\begin{split}
\rho^{\rm tot}_{\alpha \beta}(t) \equiv&\: \bra{\alpha}{} \rho^{\rm tot}(t) \ket{\beta}{} \\ 
=&\: \frac{ \mu_\alpha \mu_\beta^*}{\hbar^2}  e^{-i \omega_{\alpha \beta}t} 
\int_{0}^t  d\tau_1 \, e^{i \omega_\alpha \tau_1}   \int_{0}^t  d\tau_2 \, e^{-i \omega_\beta \tau_2} \\
\times &\lavg E\n(\tau_2)^*E\n(\tau_1) \ravg \,,
\end{split}
\end{equation}
with $\omega_{\alpha \beta} = \omega_\alpha - \omega_\beta$.

Now we require that this density matrix be representative of  thermal-light excitation,
and obtain the restrictions on the fields $E\n(t)$ and ensemble $\lavg \dots \ravg$. 
For the molecule considered in (\ref{eq:H}), interaction with thermal light would lead to the excited state
described by the density matrix $\rho^{\rm tot}$ with elements: 
\begin{equation} \label{eq:rho_known}
\begin{split}
\rho^{\rm th}_{\alpha \beta}(t) =&   \frac{\mu_\alpha \mu_\beta^*}{\hbar^2}    e^{-i \omega_{\alpha \beta}t}  
 \int_{0}^t  d\tau_1 \, e^{i \omega_\alpha \tau_1}   \int_{0}^t  d\tau_2 \, e^{-i \omega_\beta \tau_2} \\
&\times G^{(1) {\rm th}}_{ii}(\tau_1-\tau_2) \,.
\end{split}
\end{equation}
where the first-order correlation function between the $i$ and $j$  Cartesian components of the field, for thermal light, is \cite{MandelWolfBook13}:
\begin{equation}\label{eq:g1_th}
G_{ij}^{(1) \rm th}(\tau) ={}\delta_{ij} \frac{\hbar }{6 \pi^2 \epsilon_0 c^3} \int_0^\infty d\omega \, \omega^3 \, n(\omega) e^{-i \omega \tau} ,
\end{equation}
with $n(\omega)=(e^{ \hbar \omega/(k_B T)} -1)^{-1}$ being the average photon number at temperature $T$ and $k_B$ the Boltzmann's constant.

Requiring that (\ref{eq:rho_tot})  describes the same molecular dynamics as that after thermal-light excitation implies a matching of 
first-order correlation functions according to: 
\begin{equation} \label{eq:match_G1}
\lavg E\n(\tau_2)^* E\n(\tau_1) \ravg = G^{(1) {\rm th}}_{ii}(\tau_1-\tau_2) .
\end{equation}

To calculate the ensemble-averaged auto-correlation function of the field, we use the Fourier representation for the individual realizations \footnote{Here, we restrict ourselves to a one-dimensional field. A full three-dimensional analysis has been presented by Chenu \emph{et al.} \cite{Chenu2014c}).}:  
\begin{equation}\label{eq:E_ind_def}
E\n(t) =\int_0^{\infty} d\omega \: \tilde{E}\n(\omega) \: e^{-i \omega t},
\end{equation}
 where $\tilde{E}\n(\omega)$ denotes the Fourier components. 
We choose all individual fields with  identical spectral distribution $|\tilde{E}\n(\omega)|$ and only allow differences in the phase factor. The Fourier coefficients  can therefore be written as the product of a real function $f(\omega)$ defining the spectral distribution (independent of $\gamma$) and a phase factor $\phi\n(\omega)$ (dependent on $\gamma$):
\begin{equation} \label{eq:E_omega}
\tilde{E}\n(\omega)  = f(\omega) \, e^{i \phi\n(\omega)}.
\end{equation}
The field auto-correlation function averaged over the ensemble of realizations then becomes:  
 \begin{eqnarray}\label{eq:g1_def}
 \lavg E\n(\tau_2)^* E\n(\tau_1) \ravg &=& 
 \Int_0^{\infty} \Int_0^{\infty}  d\omega'  d\omega f(\omega') f(\omega) e^{i (\omega' \tau_2-\omega \tau_1)} \notag  \\
&&\times  \lavg e^{i \left(\phi\n(\omega)-\phi\n(\omega')\right)}  \ravg . 
\end{eqnarray}
We choose the phase factor with a linear dependence on frequency, and interpret the ensemble averaging as a continuous sum over the phases, i.e. 
\begin{subequations} \label{eq:phase}
\begin{align}
\phi \n(\omega) &\rightarrow \gamma \: \omega \label{eq:phase_def}\\
\lavg \bullet \ravg &\rightarrow \int_{-\infty}^{\infty} d\gamma \: \bullet, \label{eq:avg_def}
\end{align}
\end{subequations} 
so that the last line in (\ref{eq:g1_def}) yields a delta function and reduces the two frequency integrals to a single one.
Asking that (\ref{eq:g1_def}) matches the thermal-light first-order correlation function (\ref{eq:g1_th})
 restricts the lineshape $f(\omega)$ so that the electric field for each individual realization (\ref{eq:E_ind_def}) becomes: 
\begin{equation}\label{eq:E_ind}
E\n(t) = \sqrt{\frac{\hbar }{12 \pi^3 \epsilon_0 c^3} } \int_0^{\infty} d\omega  \sqrt{\omega^3 n(\omega)} e^{-i \omega (t-\gamma)}  .
 \end{equation}

Using this field as the individual field realizations, the individual excitation events (\ref{eq:psi_alpha_def}) become
\begin{equation}\label{eq:psi_ind}
\begin{split}
\ket{\psi\n(t)}{}  = K \sum_{\alpha=1}^{N_\alpha} \mu_\alpha   C_{\alpha}(t,\gamma)   e^{-i \omega_\alpha t} \ket{\alpha}{}, 
\end{split}
\end{equation}
where the complex constant $K$ and coefficients $C_{\alpha}(t,\gamma)$ are detailed in the Supplementary Information (SI). 
The state $\ket{\psi\n(t)}{}$ represents a contribution to the total density matrix following excitation from one single realization of the field. 
The total density matrix  (\ref{eq:rho_tot_def}) is then obtained by averaging over the ensemble according to our definition (\ref{eq:avg_def}). 

\begin{figure}
\includegraphics{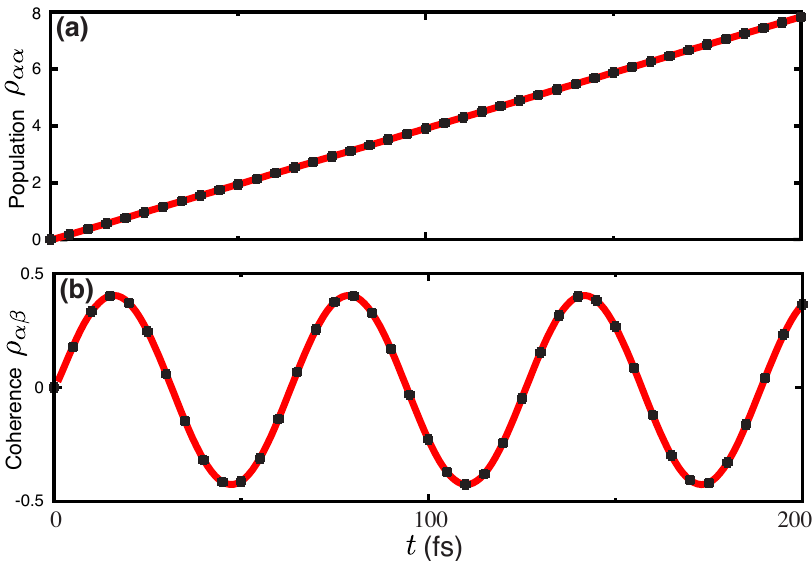}
\caption{Sample of the molecular (a) population $\rho_{\alpha \alpha}$ and (b) coherence $\rho_{\alpha \beta}$: the total response, calculated from the sum over the single excitation events (\ref{eq:rho_tot_sol})---$\rho^{\rm tot}(t)$ lines, matches the response to thermal-light excitation  (\ref{eq:rho_known})---$\rho^{\rm th}(t)$ dots. Note the different scales and amplitudes of coherence \textit{vs.} population. Parameters used are representative of a photosynthetic light-harvesting complex: $\epsilon_\alpha=1$~eV, $\epsilon_\beta=1.1$~eV, $\mu_\alpha = \mu_\beta=1$~e.\AA. Convergence of $\rho^{\rm tot}(t)$ is obtained using $20~t$ realizations, where $t$ is the time in femtoseconds.  \label{fig:rho_tot}}
\end{figure}

 It is easy to analytically show that, with our definition of the field realizations (\ref{eq:E_ind}) and using  (\ref{eq:avg_def}), the density matrix  (\ref{eq:rho_tot_def}) built from the ensemble of individual excitation events recovers that obtained directly from thermal light excitation (\ref{eq:rho_known}), i.e. 
\begin{equation}\label{eq:rho_equal}
\rho^{\rm tot}(t) = \rho^{\rm th}(t).
\end{equation}
Specifically, (\ref{eq:rho_tot}) becomes
\begin{equation} \label{eq:rho_tot_sol}
\rho_{\alpha \beta}^{\rm tot}(t) =K^2 \mu_\alpha   \mu_\beta^*    e^{-i \omega_{\alpha\beta} t} \int_{-\infty}^{+\infty} d\gamma \:  C_{\alpha}(t,\gamma) C^*_{\beta}(t,\gamma)   .
\end{equation}
Sample responses are presented in Fig. (\ref{fig:rho_tot}), showing the perfect match and equivalence between methods. 
As in \cite{Sadeq2014a, Tscherbul2014b,Olsina2014a}, where the incoherent radiation is modeled as white-noise, we find that sudden turn-on incoherent excitation can still lead to coherent excitation, with the amplitude of the coherence inversely proportional to the energy splitting between eigenstates. In our approach, the  coherence survives destructive interference between the individual excitation events with primary contribution from realizations with $0\leq \gamma \leq t$. Although non-zero, the coherences are of constant amplitude and quickly, i.e. after a few femtoseconds, become negligible compared to linearly growing populations. 

\begin{figure}
 \includegraphics{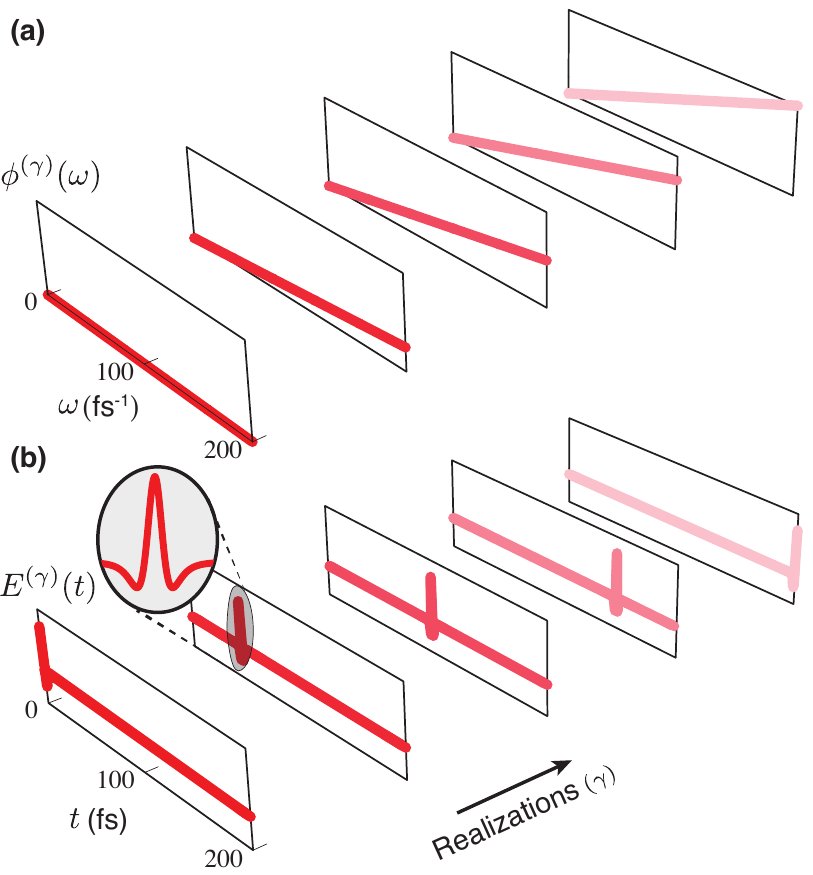}
\caption{Illustration of (a) the phase and (b) the real part of the electric field's positive frequency component (\ref{eq:E_ind}) representing a sample of field realizations---for $\gamma=\{0; 50; 100; 150; 200\}$. All fields have the same spectral distribution and coherence time, matching that of blackbody radiation. The linear phase (\ref{eq:phase_def}) results in localized pulses, with a duration of $\sim$5~fs, and acts such that the field realizations only differ from one another by a translation in time. \label{fig:E_ind}}
\end{figure}

\emph{Interpretations} -- 
By construction, the spectral density of each field realization (\ref{eq:E_ind}) is identical to that of  Planck's radiation law because of the matched correlation functions (\ref{eq:match_G1}).   Hence the field realizations's coherence time matches that of thermal light (about 1.3 fs)---since, we recall, the coherence time is defined as the inverse of the spectral density FWHM and therefore independent of the phase factor, \emph{cf.} (\ref{eq:E_omega}).
As seen in Fig. (\ref{fig:E_ind}), the  linear dependence of the phase on frequency (\ref{eq:phase_def}) localizes the individual realizations, forming pulses. In addition, the $\lambda \omega$ phase factor acts such that the field realizations only differ from one another by a time translation---also evident from the time representation in Fig. (\ref{fig:E_ind}). 
Thus, the averaging process over all realizations (\ref{eq:avg_def}) becomes equivalent to averaging over all times,
 what recovers the classical time average.

\begin{figure}
\includegraphics{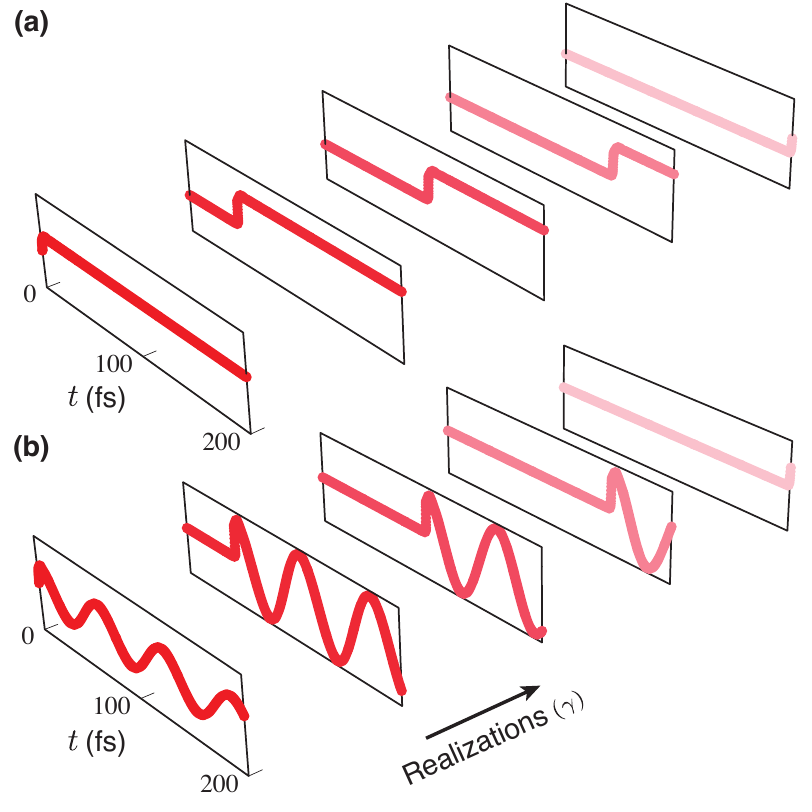}
\caption{Illustration of the individual contributions to the total molecular response, each correspond to excitation from the particular field realizations presented in Fig. (\ref{fig:E_ind}): (a) population $|\braket{\alpha}{\psi\n_{\alpha}(t)}|^2$ and (b)  coherence $\braket{\alpha}{\psi\n_{\alpha}(t)} \braket{\psi\n_{\beta}(t)}{\beta}$.  \label{fig:rho_ind}}
\end{figure}

Fig. (\ref{fig:rho_ind}) shows the individual  excitation events (\ref{eq:psi_ind}) corresponding to the particular field realizations sampled in Fig. (\ref{fig:E_ind}).
Each event corresponds to a response from an impulsive driving force, i.e. with coherent evolution and constant population after excitation \cite{Chenu2014a}. 
Although the field realizations are all identical to within a time factor,  the individual molecular responses are more complex. Specifically, the  field realizations differ by a phase factor, reflected in a time shift of the 
 individual diagonal elements contributing to the total populations $\rho_{\alpha \alpha}^{\rm tot}(t)$. 
On the other hand, coherence-wise, the individual molecular contributions $\rho_{\alpha \beta}^{\rm tot}(t)$ present, in addition to the time translation, variations in amplitude. Consequently, although the individual molecular responses are not phase related (because of different onsets),  for sudden turn-on, the total contribution can still exhibit some non-zero  oscillatory coherences surviving destructive interferences.

\emph{Discussion} -- 
The pure states (\ref{eq:psi_ind}) correspond to excitation by the field realizations (\ref{eq:E_ind}) and it is possible to build, from these individual responses, a molecular density matrix equivalent to that resulting from thermal-light excitation. Experimentally, this would require recording the molecular response after excitation at all times and averaging over the ensemble of these individual responses. 

It is noteworthy that the ensemble average can be performed in many ways. 
Similar results fulfilling (\ref{eq:rho_equal}) can be obtained with individual realizations that differ from (\ref{eq:E_ind})---e.g. non-pulsed light could be used instead. More generally, the only requirement to achieve the thermal result is to match the thermal-light first-order correlation function, which requires: (i)  matching the spectral distribution (and hence the coherence time), and (ii) that the ensemble of field realizations fulfills 
 $ \lavg e^{i \left(\phi\n(\omega')-\phi\n(\omega)\right)}  \ravg  \propto \delta(\omega- \omega')$.
These conditions  will reproduce the stochastic character of the first-order correlation of thermal-light.
The latter requirement resembles the condition that the Fourier components of a stationary random process belonging to different frequencies are uncorrelated \cite{MandelWolfBook2}. 
While the Fourier components of the derived fields (\ref{eq:E_ind}) are correlated at the level of the individual realizations, this correlation is lost upon averaging over an infinity of independent realizations.
This is in accordance with the fact that, for a stationary and ergodic ensemble of random functions, one can replace the auto-correlation function defined by the time average by that defined as an ensemble average \cite{MandelWolfBook4}. 
Our results demonstrate that this  not only holds at the level of the field, but also applies to describe linear excitation of molecules.

We emphasize that our results apply to \emph{linear} interaction, relevant to natural incoherent light. In general, retrieving the state of thermal light (i.e. its full density matrix and match all its properties) from an ensemble of single broadband coherent states (representative of pulses) is not possible and already fails for the simplest second-order correlation describing broadband delayed excitation \cite{Chenu2014b}. Further, we emphasize that the individual realizations have no physical meaning. It is only the ensemble average that relates to the natural incoherent excitation process. 

In conclusion, the implications of our results are two-fold for the modelling of linear light-matter interaction: conceptually, we showed that linear excitation by thermal light can be represented by an ensemble average of deterministic pulse realizations, with no random character---all effects of incoherence are  recast in the post excitation averaging. Practically, we provide, with Eq. (\ref{eq:psi_ind}), an alternative description of the molecular excitation that allows one to study the dynamics of a molecule at the wave function level, which presents great simplifications over the conventional density matrix formalism. 
Further, the results motivate experimental studies of natural incoherent excitation using pulsed laser techniques.
As such, our findings are expected to facilitate the study of the molecular dynamics after thermal-light excitation, such as photosynthetic light-harvesting complexes under natural excitation conditions. 

\begin{acknowledgements}
We thank Dr. T. V. Tscherbul for comments on the manuscript. A.C. acknowledges funding from the Swiss National Science Foundation. P.B.'s research was supported by AFOSR under contract number FA 9550-13-1-0005 and NSERC. 
\end{acknowledgements}

\clearpage

\appendix
\onecolumngrid

\section{Dynamics of a molecular system excited by one realization of the field}
Here we provide details of the wave function dynamics after excitation by one realization of the field (\ref{eq:psi_alpha_def}). 
Taking, as exciting field, the individual realizations derived in (\ref{eq:E_ind}) yields: 
\begin{eqnarray}\label{eq:psi_app}
\ket{\psi\n(t)}{}  = - \frac{i}{\hbar} \sqrt{\frac{\hbar }{12 \pi^3 \epsilon_0 c^3}  }  \: \sum_{\alpha}^{N_\alpha}\mu_\alpha e^{-i \omega_\alpha t} \int_{0}^t  d\tau \, e^{i \omega_\alpha \tau}%
p_1(\tau-\gamma) \ket{\alpha}{}\, .
\end{eqnarray}
Here, we have assumed the field to lie along the orientation of the transition dipole moment $\uv{n}_\alpha$, and we have defined
\begin{equation*}
p_\nu(\tau)\equiv  \int_0^\infty  d\omega \: \omega^\nu \sqrt{\omega  \: n(\omega)} e^{-i \omega \tau} .
\end{equation*}
Expanding  $\sqrt{n(\omega)}\equiv e^{-\nicefrac{ \hbar \omega}{2 k_B T}} (1-e^{- \nicefrac{\hbar \omega}{k_B T}})^{-1/2}$ in terms of power of exponentials yields to a series of terms for which the integral can be evaluated, i.e. 
\begin{equation*}
\begin{split}
p_0(\tau) =& \sum_{j=0}^\infty \binom{-\frac{1}{2}}{j} (-1)^j \int_0^\infty  d\omega \: \sqrt{\omega} \:  e^{-[(\nicefrac{1}{2}+ j)\nicefrac{ \hbar}{ k_B T}  +i \tau]\omega } \\
=& \sum_{j=0}^\infty \binom{-\frac{1}{2}}{j}  \frac{(-1)^j \sqrt{\pi}}{2 [ (\nicefrac{1}{2}+ j)\nicefrac{ \hbar}{ k_B T}   +i \tau]^{3/2}}\, .
\end{split}
\end{equation*}
Note that $p_\nu(t) = (-1/ i)^\nu \frac{\partial^\nu}{\partial t ^\nu} p_0(t)$. We therefore take the time derivative to obtain  
\begin{equation*}
\begin{split}
p_{1}(\tau) =& \frac{3 \sqrt{\pi}}{4}\sum_{j=0}^\infty \binom{-\frac{1}{2}}{j} (-1)^j \left[(\nicefrac{1}{2}+ j)\nicefrac{ \hbar}{ k_B T}  +i \tau\right]^{-5/2}. 
\end{split}
\end{equation*}
We add the integration in time and define: 
\begin{equation}\label{eq:C_def}
\begin{split}
 C_{\alpha,j}(t,\gamma)  \equiv& \sum_{j=0}^\infty  \binom{-\frac{1}{2}}{j} (-1)^j  \, \int_{0}^t  d\tau \, \frac{e^{i \omega_\alpha \tau}}{\left[(\nicefrac{1}{2}+ j)\nicefrac{ \hbar}{ k_B T}  +i (\tau-\gamma)\right]^{5/2} }  \\
    =&\sum_{j=0}^\infty  \binom{-\frac{1}{2}}{j} (-1)^j  \,\left[\frac{2}{3} i \left[ J_{\alpha,j}(t,\gamma)- J_{\alpha,j}(0,\gamma)\right]%
    -\frac{8}{3} i \omega_\alpha^{3/2} %
     \left(e^{i t \omega_\alpha}  F\sqrt{\omega_\alpha  h_{j}(t,\gamma)}- F\sqrt{\omega_\alpha  h_{j}(0,\gamma)}\right) \right]
 \end{split}
\end{equation}
with $F(x) \equiv e^{-x^2} \int_0^x e^{y^2} dy$ is the Dawson function and:
\begin{eqnarray*}
h_{j}(t,\gamma)&\equiv&  (\nicefrac{1}{2} + j) \nicefrac{\hbar}{k_B T} +i (t-\gamma)\\
J_{\alpha,j}(t,\gamma) &=&  e^{i t \omega_\alpha} \frac{2 \omega_\alpha h_{j}(t,\gamma) + 1}{h_{j}(t,\gamma)^{3/2}} 
\end{eqnarray*}
Although (\ref{eq:C_def}) involves an infinite number of terms, the sum converges quickly and, at room temperature, results are typically within $2\%$ accuracy accounting for the first three terms only.

Taking 
\begin{equation}
K=- i  \sqrt{\frac{ 3 }{64 \pi^2 \epsilon_0 \hbar c^3}},
\end{equation} 
the dynamics of system (\ref{eq:psi_app}) reduces to the expression given in the main text (\ref{eq:psi_ind}).

\end{document}